\documentclass[%
 reprint,
nofootinbib,
 amsmath,amssymb,
 aps,
]{revtex4-1}
\usepackage{url}
\usepackage{hyperref}
\usepackage{subfigure}
\usepackage{graphicx}% Include figure files
\usepackage{dcolumn}% Align table columns on decimal point
\usepackage{slashed}
\usepackage{xcolor}
\usepackage{bm}% bold math
\def\beq{\begin{equation}}
\def\eeq{\end{equation}}
\def\be{\begin{equation}}
\def\ee{\end{equation}}
\def\bea{\begin{eqnarray}}
\def\eea{\end{eqnarray}}

\def\e{\eta}

\DeclareMathOperator{\s}{\sigma}
\DeclareMathOperator{\bs}{\bar{\sigma}}

\def\g{\gamma}

\begin{document}
\title{Pure gauge theory for the gravitational spin connection}
\medskip\
\author{Stephon Alexander}%
\email[Email:]{ stephon_alexander@brown.edu}
\affiliation{Department of Physics,
Brown University, Providence, RI 02912, USA}
\author{Tucker Manton} 
\email[Email:]{ tucker_manton@brown.edu}
\affiliation{Department of Physics,
Brown University, Providence, RI 02912, USA}

\date{\today}% It is always \today, today,

\begin{abstract}
The gravitational spin connection appears in gravity as a non-Abelian gauge field for the Lorentz group $SO(3,1)$, which is non-compact. The action for General Relativity is linear in the field strength associated to the spin connection, and its equation of motion corresponds to the standard metricity constraint. Consequently, the zero-torsion spin connection is never realized as an independent degree of freedom and is determined by the vierbein field. In this work, we take a different perspective and consider a pure Yang-Mills theory for the spin connection coupled to Dirac fermions, resulting in the former being a dynamical field. After discussing various approaches towards managing the pathologies associated with non-compact gauge theories, we compute the tree-level amplitude for fermion scattering via a spin connection exchange. In contrast to integrating out torsion in the presence of fermions, the model induces a chiral four-Fermi like term that involves a right-right current interaction, which is not present in the Standard Model.
\end{abstract}

\pacs{Valid PACS appear here}% PACS, the Physics and Astronomy
\maketitle
%=================================================================================================================================
%=============================================================================================================================
\section{Introduction} 

Shortly after Dirac published his groundbreaking work on the quantum theory of the electron \cite{Dirac:1928hu}, efforts were immediately directed towards coupling spin-1/2 particles to General Relativity \cite{Weyl:1929fm}. It is now well known that the most natural setting in which to embed spin-1/2 interactions in General Relativity is Einstein-Cartan theory \cite{Cartan:1923zea}, which was in fact written down before the discovery of spin \cite{Trautman:2006fp}. The modern viewpoint can be considered in a close analogy with the Standard Model gauge theories. That is, we begin with the free Dirac Lagrangian $\mathcal{L}_{Dirac}=\Bar{\Psi}(i\slashed{\partial}-m)\Psi$ and define a local Lorentz symmetry $\Psi(x)\rightarrow\Psi'(x')=S(x)\Psi(x)$, where $S(x)$ has the exponential representation
\begin{equation}\label{LorentzSym}
    S(x)=\text{exp}\Big(-\frac{i}{2}J_{ab}\alpha^{ab}(x)\Big).
\end{equation}
Here, $J_{ab}=\frac{i}{4}[\gamma_a,\gamma_b]$ is the generator of the $SO(3,1)$ algebra whose commutator satisfies $[J_{ab},J_{cd}]=2if_{abcdef}J^{ef}$ and $\gamma^a$ are the Dirac matrices. The structure constants $f_{abcdef}=f_{[ab][cd][ef]}$ can be expanded in terms of the Minkowski metric. Invariance of the theory is enforced by promoting the derivative in the Dirac Lagrangian to
\begin{equation}
    D_\mu=\partial_\mu -\frac{ig}{2}J_{ab}A^{ab}_{\mu},
\end{equation}
where $A^{ab}_\mu$ is the spin connection whose transformation $J_{ab}A^{ab}_\mu\rightarrow S(J_{ab}A^{ab}_\mu)S^{-1}-\frac{1}{2g}J_{ab}\partial_\mu \alpha^{ab}(x)$ preserves the Lorentz symmetry (\ref{LorentzSym}. A field strength for the spin connection is defined in the standard way as $[D_\mu,D_\nu]=-\frac{ig}{2}J_{ab}F^{ab}_{\mu\nu},$ from which we obtain
\begin{equation}\label{fieldstrength}
    F_{\mu\nu}^{ab}=\partial_{[\mu}A^{ab}_{\nu]}+gf^{ab}_{ \ \ cdef}A^{cd}_\mu A^{ef}_\nu.
\end{equation}

In order to make contact with gravity, we introduce the spacetime vierbein $e^a_{ \ \mu}$, which relates to the metric as $g_{\mu\nu}=e^a_{ \ \mu}e^b_{ \ \nu}\eta_{ab},$ where $\eta_{ab}=\text{diag}(-1,1,1,1)$\footnote{We follow notation where the internal $SO(3,1)$ (or frame) indices are $\{a,b,...\}$ while the spacetime indices are $\{\mu,\nu,...\},$ and our antisymmetrization convention includes no factor of 2, $i.e.$ $X_{[a}Y_{b]}=X_aY_b-X_bY_a$.}. As usual, the vierbeins connect frame fields to spacetime fields, \textit{i.e} $V_\mu = V_a e^a_{ \ \mu}$ for a spacetime vector $V_\mu$. We therefore arrive at the following Lagrangian for a Dirac spinor over curved space,
\begin{equation}\label{LDirac}
    \mathcal{L}_{\text{Dirac}}=\bar{\Psi}\Big(ie^\mu_{ \ a}\gamma^a \big[\partial_\mu -\frac{ig}{2}J_{bc}A^{bc}_{\mu}\big]-m\Big)\Psi.
\end{equation}
Next consider the action
\begin{equation}\label{GRaction}
    S=\int d^4x ee^\mu_{ \ a}e^\nu_{ \ b}F^{ab}_{\mu\nu}(A,\partial A),
\end{equation}
where $e=\text{det}(e^a_{ \ \mu})$ and $F^{ab}_{\mu\nu}$ is given by (\ref{fieldstrength}). The equation of motion associated to the spin connection $A^{ab}_\mu$ is proportional to the covariant derivative of the vierbein, 
\begin{equation}\label{nablae}
    \nabla_\mu e^a_{ \ \nu}=\partial_\mu e^a_{ \ \nu}-\Gamma_{\mu\nu}^\lambda e^a_{ \ \lambda}+gA^a_{b\mu}e^b_{ \ \nu},
\end{equation}
where $\Gamma^\lambda_{\mu\nu}$ are the metric compatible Christoffel symbols. By setting (\ref{nablae}) equal to zero, we can trivially solve for the spin connection in terms of the vierbein, $A^{ab}_\mu=A^{ab}_\mu(e).$ Inserting back into the action (\ref{GRaction}), we recover standard General Relativity with the Ricci scalar given by
\begin{equation}
    R=e^\mu_{ \ a}e^\mu_{ \ b}F_{\mu\nu}^{ab}(e)
\end{equation}
We will call the vanishing of (\ref{nablae}) the \textit{metricity constraint}, which is often imposed a priori such that the spin connection is never realized as an independent degree of freedom. This is a feature of the theory being linear in the field strength in the absence of other matter fields.

Recently, in \cite{Donoghue:2016vck}, Donoghue offered an  argument for abandoning the metricity constraint and treating the spin connection as an independent degree of freedom. Consider now the theory 
\begin{equation}\label{Lagrangian}
    \mathcal{L}=-\frac{1}{4}F_{\mu\nu}^{ab}F_{ab}^{\mu\nu}+\mathcal{L}_{\text{Dirac}},
\end{equation}
where the Dirac Lagrangian is given by (\ref{LDirac}). In essence, this is just Yang-Mills coupled to fermions, albeit for the non-compact gauge group $SO(3,1).$ It was shown in \cite{Donoghue:2016vck} that for the theory (\ref{Lagrangian}), the one-loop $\beta$-function for the spin connection coupling $g$ is negative,
\begin{equation}
    \beta(g)=-\frac{22}{3}\frac{g^3}{16\pi^2},
\end{equation}
suggesting that the spin connection is confined or condensed in the infrared. It is therefore interesting to consider the form of interactions mediated by the spin connection, which is the purpose of this note. 

 There have been numerous attempts cast gravity as a gauge theory \cite{Kibble:1961ba,Poplawski:2012bw,Mansouri:1976df,Fairchild:1976we} (see also \cite{Donoghue:2017pgk} and references therein). Indeed, General Relativity can be derived by considering a global spacetime translation $x^\mu\rightarrow x^\mu+a^\mu$ and gauging such that $a^\mu\rightarrow a^\mu(x),$ analogous to promoting the Lorentz symmetry to a local operator (\ref{LorentzSym}).  The crucial difference between Yang-Mills theory and the treatment of General Relativity as a gauge theory is the curvature being quadratic in the former case and linear in the latter case. Moreover, unifying the weak force with gravity provides additional motivation to write General Relativity as a gauge theory \cite{Alexander:2011jf,Cahill:1982zf,Nesti:2007ka}. It would be fascinating if the construction outlined above serves as a road map towards a new unification approach. We will comment further on both of these points in the discussion. 

We will now proceed under the assumption that the spin connection is indeed confined or condensed and explore the properties of the theory described by (\ref{Lagrangian}), but before studying the interactions, let us briefly comment on the issue of non-compactness of $SO(3,1)$.

\section{On the non-compactness of $SO(3,1)$}\label{sec:noncompactness}
%=================================================================================================================================
%=================================================================================================================================
It is useful to take capital Roman letters $A,B,...$ as fundamental indices in order to discuss an arbitrary, non-Abelian Lie group $G$. For $SO(3,1)$, the $\{A,B,...\}$ indices are composed of pairs of internal indices $\{a,b,...\}$. We denote the Killing form as $\mathcal{K}_{AB}$, which is the internal metric for the gauge group. A simple way to examine compactness is to write out the the Yang-Mills Lagrangian,
\begin{equation}\label{GenYMLagrangian}
    \mathcal{L}= -\frac{1}{4}\mathcal{K}_{AB}F^A_{\mu\nu}F^B_{\alpha\beta}g^{\mu\alpha}g^{\nu\beta},
\end{equation}
and consider the quadratic form 
\begin{equation}
    Q=\mathcal{K}_{AB}u^Au^B
\end{equation}
for arbitrary $u\neq 0\in G$. A gauge group is compact if $Q>0$ for all nonzero group elements. For example in $SU(N),$ $\mathcal{K}_{AB}=\delta_{AB}$, which is clearly positive definite. This is analogous to the Hamiltonian density being bounded from below. For the Lorentz group $SO(3,1)$, the Killing form is $\mathcal{K}_{AB}\rightarrow\mathcal{K}_{abcd}=\eta_{ac}\eta_{bd}$. Since the Minkowski metric is not positive definite, the quadratic form $Q=\mathcal{K}_{abcd}u^{ab}u^{cd}$ can be negative or zero even when $u\neq 0.$ This implies that certain quantum states of the spin connection will have a negative norm and are thus unhealthy. However, there are a few ways to handle the potential pathology. 

In \cite{Margolin:1990wt}, the authors showed that in non-compact sigma models based on $SL(2,\mathbb{C})$, there is an inherent superselection rule that results in a zero overlap between the healthy and unhealthy (negative norm) states. Then in \cite{Margolin:1992rg}, the same authors extended the analysis to arbitrary non-compact gauge groups. The argument leans heavily on BRST symmetry \cite{Becchi:1975nq}, where in addition to (\ref{GenYMLagrangian}), the theory includes ghost and anti-ghost fields $c^A,\bar{c}^B$, as well as the Nakanishi-Lautrup field $B^A$ \cite{Nakanishi:1966zz}, the latter serving the purpose of a generalized gauge fixing procedure. The authors exploit a Cartan involution $\mathcal{D}$ on $G$ defined as $\mathcal{D}(X^A)=-\mathcal{K}^A_BX^B$, where $X^A$ is any field (including the field strength). Using the BRST transformations for $\{A^A_\mu,c^A,\bar{c}^A,B^A\}$ \cite{Fuster:2005eg} it is straightforward to show that $\mathcal{D}$ commutes with the BRST charge $Q_{brst}$, which defines the space of physical asymptotic states $\mathcal{V}_{phys}$ in the standard way, $\mathcal{V}_{phys}=\{|v\rangle\in \mathcal{V}: \ Q_{brst}|v\rangle =0\}$\footnote{An explicit expression for the BRST charge is $Q_{brst}=i\mathcal{K}_{AB}\int d^3k[\hat{b}^{A,\dagger}(k)\hat{c}^B(k)-\hat{c}^{A,\dagger}(k)\hat{b}^B(k)]$, where $\hat{c}^\dagger,\hat{b}^\dagger,\hat{c},\hat{b}$ are the creation and annihilation operators for the fields $c^A$ and $B^A$. }, where $\mathcal{V}$ is the total space of asymptotic states. Then consider the subspace $\mathcal{H}\subset\mathcal{V}$ generated by the transverse modes of the gauge fields, which can be decomposed as $\mathcal{H}=\mathcal{H}^{+}\oplus\mathcal{H}^{-}$, where $\mathcal{H}^{+}$ is characterized by the asymptotic operators with $\mathcal{K}_{AA}=+1$ while $\mathcal{H}^{-}$ is characterized by the operators with $\mathcal{K}_{AA}=-1.$  The subspaces $\mathcal{H}^{\pm}$ satisfy $\mathcal{D}(\mathcal{H}^{\pm})=\pm 1.$ Defining the physical $S$-matrix as $S_{phys}=P^{\dagger}_{\mathcal{H}}SP^{\dagger}_{\mathcal{H}}$ by projecting into $\mathcal{H}$ from the full $S$-matrix $S$, we have that $[\mathcal{D},S_{phys}]=0$. Therefore $\mathcal{D}$ defines the superselection rule between the states $|\alpha\rangle\in\mathcal{H}^{+}$ and $|\beta\rangle\in\mathcal{H}^{-}$ such that $\langle \alpha|\beta\rangle=0$, implying there is no negative probabilities. Thus, even though there are unhealthy states in the spectrum, there is a consistent procedure to project out the Hilbert space for the physical states and construct a well defined, physical $S$-matrix.

A slightly different approach was studied in \cite{Cahill:1978ps,Cahill:1979qt,Cahill:1981rq}, where the author considers the non-compact gauge groups $GL(N,\mathbb{R})$ and $GL(N,\mathbb{C})$. The Killing form is promoted to a dynamical field and is given a gauge covariant derivative (see also \cite{Alexander:2016uli} for a similar idea implemented in collider phenomenology). It then has a contribution to the Hamiltonian (density) that conspires to bound the system from below. The field equation for the Killing form is not present when the gauge group is compact. The field content of the theory is a set of massless vector mesons associated with the subgroup $U(N)\subset GL(N,\mathbb{C})$ and massive vector mesons associated with the non-compact part of $GL(N,\mathbb{C})$. For the latter, the longitudinal modes are supplied by real scalar fields that make up the components of the Killing metric (see \cite{Cahill:1979qt} for an explicit illustration of these features for the example of $GL(1,\mathbb{C})$).

On the other hand, one can consider making a gauge choice reminiscent of the temporal gauge in QED, by setting $A^{a=0,b}_\mu=0$. This removes half of the components of the spin connection and has the effect of forcing the quadratic form associated to $\mathcal{K}_{AB}$ to be positive definite. To see this, note that the components of the spin connection can be imagined as a 4x4, antisymmetric matrix of four-vectors with free index $\mu.$ The gauge fixing condition $A^{a=0,b}_\mu=0$ deletes the first row of the matrix, and the Killing form only acts non-trivially on `spatial' $SO(3,1)$ internal indices. In essence, this implies $\mathcal{K}_{abcd}\rightarrow \delta_{il}\delta_{jm}$, where $i,j,...$ take on values $\{1,2,3\}$. 

Each of these - the BRST approaches in \cite{Margolin:1990wt,Margolin:1992rg}, the dynamical approach in \cite{Cahill:1978ps,Cahill:1979qt,Cahill:1981rq}, and the gauge fixing approach - all share the common thread of projecting the healthy states into a maximal compact subgroup $H\subset G.$ Explicitly, `temporal' gauge fixing has the effect of gauging away the boosts $J^{0i}$ and we are left with the rotations of $O(3)\subset SO(3,1)$. It should not be a surprise that the remaining components of the spin connection correspond to healthy states, as the rotation matrices $J^{ij}$ are unitary, $J^{ij\dagger}=J^{ij},$ while the boosts are not, $J^{0i\dagger}=-J^{0i}$.

Importantly, the issue of negative norm states is only critical when considering the spin connection being on an external leg of a given process. This arises in, for example, fermion annihilation into two spin connections, $\Psi\Psi\rightarrow AA,$ or in a fermion scattering process involving any number of spin connection loops, where a unitarity cut will produce a diagram analogous to the $\Psi\Psi\rightarrow AA.$ However, we can consider the possibility that instead of the asymptotic vacuum being devoid of any particle states, it is occupied by some sort of condensate, such as the ghost condensate of \cite{Arkani-Hamed:2003pdi}. In fact, the ghost condensate is known to break Lorentz transformations down to just spatial rotations, which is precisely the setting in which the S-matrix for the asymptotic spin connection states is unitary.

In the next section, we focus solely on the tree level spin connection exchange, and leave the delicate treatment of the external spin connection to future explorations.

%=================================================================================================================================
%=================================================================================================================================
\section{Interactions}
%=================================================================================================================================
%=================================================================================================================================
The interaction vertices are readily obtained from the Lagrangian (\ref{Lagrangian}). The spin connection self interactions are of the form
\begin{equation}
    \begin{split}
        \text{three point}& \ \ \ \sim  \ \ \ g\partial_\mu A_\nu^{ab}f_{ab}^{ \ \ \ cdef}A^\mu_{cd}A^\nu_{ef}, \\
        \text{four point}& \ \ \ \sim \ \ \ g^2f^{ab}_{ \ \ \ cdef}A^{cd}_\mu A^{ef}_\nu f_{ab}^{ \ \ \ ghij}A^\mu_{gh}A^\nu_{ij},
    \end{split}
\end{equation}
while the fermion interaction is
\begin{equation}\label{3pt}
    \begin{split}
        \text{three point}& \ \ \ \sim \ \ \ g\bar{\Psi}\gamma^ae_a^{ \ \mu}J_{cd}A^{cd}_\mu\Psi.
    \end{split}
\end{equation}

\begin{figure}
    \centering
    \includegraphics{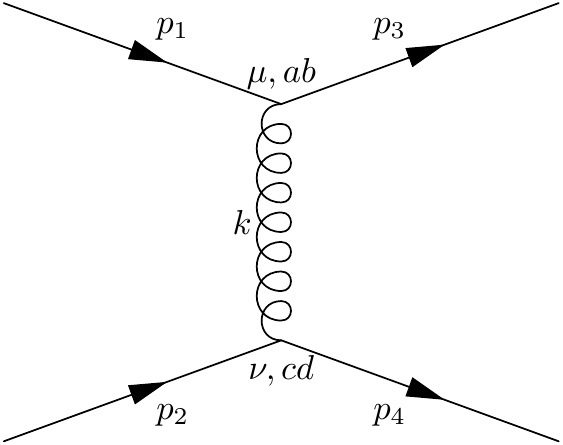}
    \caption{$t$-channel exchange}
    \label{tchannel}
\end{figure}

We will now calculate the tree-level $t$-channel exchange of a spin connection between two fermions in helicity eigenstates, where (\ref{3pt}) is the only relevant vertex. Let the incoming fermions have momenta $p_1$ and $p_2$, with outgoing momenta $p_3$ and $p_4$, with $k$ being the momentum exchange across the propagator (Fig. \ref{tchannel}). Standard application of the Feynman rules yields

\begin{equation}\label{D1}
   \begin{split}
      i\mathcal{M}_t=   &\Bigg(-\frac{ig}{2}\bar{\Psi}(p_3)e^\mu_f\gamma^f J^{ab}\Psi(p_1)\Bigg)D_{\mu\nu,abcd}(k) \\        
        & \ \ \ \ \ \ \ \ \ \times \Bigg(-\frac{ig}{2}\bar{\Psi}(p_4)e^\nu_g\gamma^g J^{cd}\Psi(p_2)\Bigg),
    \end{split}
\end{equation}
where $D^{abcd}_{\mu\nu}$ is the spin connection propagator and will be defined shortly. Note that we can pull the vierbein contractions through to the propagator. And using $J^{ab}=\frac{i}{4}[\g^a,\g^b]$, we can simplify (\ref{D1}) to 
\begin{equation}\label{D2}
\begin{split} 
  i\mathcal{M}_t=&  \frac{g^2}{64}\Bigg(\bar{\Psi}(p_3)\gamma^f[\gamma^a,\gamma^b]\Psi(p_1)\Bigg)e^\mu_f D_{\mu\nu,abcd}(k)e^\nu_g\\
  & \ \ \ \ \ \ \ \ \ \ \ \times \Bigg(\bar{\Psi}(p_4)\gamma^g[\gamma^c,\gamma^d]\Psi(p_2)\Bigg).
  \end{split}
\end{equation}
We next look to rewrite the product of Dirac matrices, which is accomplished utilizing the relation
\begin{equation}
\g^f\g^{[a}\g^{b]}=i\epsilon^{fabh}\g_h\g^5+2\eta^{f[a}\g^{b]}.
\end{equation}
We have
\begin{equation}\label{D3}
\begin{split} 
 & i\mathcal{M}_t= \\ &  \frac{g^2}{64}\Bigg(\bar{\Psi}(p_3)\big(i\epsilon^{fabh}\g_h\g^5+2\eta^{f[a}\g^{b]}\big)\Psi(p_1)\Bigg)e^\mu_f D_{\mu\nu,abcd}(k)e^\nu_g \\
  & \ \ \ \ \ \ \ \ \ \ \ \times \Bigg(\bar{\Psi}(p_4)\big(i\epsilon^{gcdi}\g_i\g^5+2\eta^{g[c}\g^{d]}\big)\Psi(p_2)\Bigg).
  \end{split}
\end{equation}
At this step, recall that the vector and axial currents are given by
\begin{equation}
    J^a_V=\bar{\Psi}\g^a\Psi, \ \ \ \ \ J^a_A=\bar{\Psi}\g^a\g^5\Psi.
\end{equation}
We can thus rewrite (\ref{D3}) as
\begin{equation}\label{D4}
\begin{split}
i\mathcal{M}_t=&    \frac{g^2}{64}\Big(i\epsilon^{fabh}J_{A,h}+2\eta^{f[a}J_V^{b]}\Big)e^\mu_f D_{\mu\nu,abcd}(k)e^\nu_g \\
& \ \ \ \ \ \ \ \ \ \times \Big(i\epsilon^{gcdi}J_{A,i}+2\eta^{g[c}J_V^{d]}\Big).
\end{split}
\end{equation}
This result shows that the particular form of the four-Fermi interaction is contingent on the index structure of the propagator $D_{\mu\nu,abcd}$. To obtain the tree level propagator, we consider the derivative terms in the spin connection Lagrangian including a gauge-fixing piece, 
\begin{equation}
\begin{split} 
   & \mathcal{L}\supset \eta_{ac}\eta_{bd}g^{\mu\alpha}g^{\nu\beta} \\
   & \ \ \ \ \times \Big(-\frac{1}{4} \partial_{[\mu}A^{ab}_{\nu]}\partial_{[\alpha}A_{\beta]}^{cd}-\frac{1}{2\xi}(\partial_\alpha A_\mu^{ab})(\partial_\beta  A_\nu^{cd})\Big)
   \end{split}
\end{equation}
The inverse of the full derivative operator is the propagator
\begin{equation}\label{prop}
        iD^{abcd}_{\mu\nu}(k)=\frac{-i\Big(g_{\mu\nu}-(1-\xi)\frac{k_\mu k_\nu }{k^2}\Big)}{k^2+i\epsilon}\frac{1}{2}1_{4}\big(\e^{ac}\e^{bd}-\e^{ad}\e^{bc}\big).
\end{equation}
Taking the Feynman gauge $\xi=1,$ (\ref{D4}) simplifies to 
\begin{equation}\label{D5}
\begin{split}
&i\mathcal{M}_t= \\&    -\frac{g^2}{64}\Big(i\epsilon^{fabh}J_{A,h}+2\eta^{f[a}J_V^{b]}\Big)\frac{\eta_{fg}\frac{1}{2}1_4(\eta_{ac}\eta_{bd}-\eta_{ad}\eta_{bc})}{k^2+i\epsilon}\\
& \ \ \ \ \ \ \ \ \ \ \ \times \Big(i\epsilon^{gcdi}J_{A,i}+2\eta^{g[c}J_V^{d]}\Big).
\end{split}
\end{equation}
Note the interesting observation that by virtue of the vierbeins appearing in the interaction, the full spacetime metric $g_{\mu\nu}$ is projected to the frame metric as $e^\mu_f e^\nu_g g_{\mu\nu}=\eta_{fg}.$ 

It is straightforward to contract out the remaining frame indices, and dropping the $i\epsilon$, we find the simple result
\begin{equation}
  i\mathcal{M}_t=  g^2\frac{3}{32k^2}\Big(J_A^{13}\cdot J_A^{24}+J_V^{13}\cdot J_V^{24}\Big),
\end{equation}
where the $\{1234\}$ superscripts denote the momenta dependence. If we suppose that this expression is valid solely below some momentum scale $k_c$, then we may write the four-Fermi interaction
\begin{equation}\label{4fermi1}
  i\mathcal{M}_t=  \tilde{G}_F\Big(J_A^{13}\cdot J_A^{24}+J_V^{13}\cdot J_V^{24}\Big),
\end{equation}
defining a `Fermi constant' $\tilde{G}_F=\frac{3g^2}{32k_c^2}$. Finally, recall that the axial and vector currents are related to the helicity eigenstate currents $J_L^a=\bar{\psi}_L\g^a\psi_L$ and $J_R^a=\bar{\psi}_R\g^a\psi_R$ by
\begin{equation}
    J_A^a=J_R^a-J_L^a, \ \ \ \ \ J_V^a=J_R^a+J_L^a.
\end{equation}
This allows us to recast (\ref{4fermi1}) as
\begin{equation}\label{4fermi2}
  i\mathcal{M}_t=  2\tilde{G}_F\Big(J_R^{13}\cdot J_R^{24}+J_L^{13}\cdot J_L^{24}\Big).
\end{equation}
There is a similar contribution to the full amplitude from a $u$-channel diagram, which simply exchanges the final state momenta, $p_3\leftrightarrow p_4.$ The result (\ref{4fermi2}) shows that there is no mixing between the helicity eigenstates in this theory, and there is a pure right-right interaction.

%=================================================================================================================================
%=================================================================================================================================
\section{Discussion}
%=================================================================================================================================
%=================================================================================================================================

In this work, we have embraced the possibility that the gravitational spin connection is condensed or confined at low energy, and considered the tree level fermion scattering process mediated by the spin connection. We believe that the theory (\ref{Lagrangian}) is well motivated following the conclusions met in \cite{Donoghue:2017pgk}. 

The theory admits a four-Fermi interaction where there is no mixing between the helicity eigenstates. This is in contrast to very similar calculations performed in \cite{Cianfrani:2015yya} and \cite{Nesti:2007ka,Alexander:2007mt,Alexander:2009uu,Perez:2005pm,Freidel:2005sn}. In \cite{Alexander:2009uu,Perez:2005pm}, the spin connection is decomposed as $A=\tilde{A}+C$, where $\tilde{A}$ is metric compatible and $C$ is related to torsion. The torsion piece is integrated out of the theory and results in a purely axial current interaction, which mixes left and right handed currents. A slightly more general approach was taken in \cite{Freidel:2005sn}, where the fermion Lagrangian contains a `non-minimal coupling parameter' $\alpha$,
\begin{equation}\label{fermions}
\mathcal{L}\sim (1-i\alpha)\bar{\Psi}\gamma^ae^\mu_a\nabla_\mu\Psi-(1+i\alpha)\overline{\nabla_\mu\Psi}\gamma^ae^\mu_a\Psi,
\end{equation}
which, after arguing that consistency with the no-torsion constraint demands $\alpha\in\mathbb{R}$, they obtain
\begin{equation}\label{a1}
\mathcal{L}_{int}\sim\frac{\gamma^2}{\gamma^2+1}\Big(J_A\cdot J_A+\frac{2\alpha}{\gamma}J_A\cdot J_V-\alpha^2J_V\cdot J_V\Big).
\end{equation}
(The result from \cite{Alexander:2009uu,Perez:2005pm} is simply the minimal coupling $\alpha=0$.) In (\ref{a1}), $\gamma$ is the Immirzi parameter of loop quantum gravity which appears, for example, in the Holst action \cite{Holst:1995pc}. Our result (\ref{4fermi1}) is nicely consistent with the discussion surrounding equation (11) of \cite{Freidel:2005sn}; (\ref{4fermi1}) is obtained from (\ref{a1}) in the limit $\{\alpha,\gamma\}\rightarrow\{\pm i,\infty\}$. In our approach, we explicitly relax the no-torsion constraint, which in the language of \cite{Freidel:2005sn}, relaxes the reality condition on $\alpha$, such that $\alpha=\pm i$ is indeed sensible. The $\gamma\rightarrow \infty$ kills off the parity violating cross term, which would moreover leave one with an imaginary Lagrangian in (\ref{a1}) if $\alpha=\pm i$. So in one sense, our result appears to emerge from a theory equivalent to that studied in \cite{Freidel:2005sn} allowing torsion along with the `non-minimally coupled' fermions (\ref{fermions}) in the limit $\{\alpha,\gamma\}\rightarrow\{\pm i,\infty\}.$

On the other hand, the authors of \cite{Cianfrani:2015yya} consider the so-called Fairchild theory \cite{Fairchild:1977wi}, which is similar to \ref{Lagrangian} except with an Einstein-Hilbert term linear in $F^{ab}_{\mu\nu}$. After linearizing about flat space $(e^a_{ \ \mu}\approx \delta^a_\mu$), decomposing the spin connection into irreducible components, and neglecting terms quadratic in the torsion, it is shown that healthy fermionic tree-level interactions decouple from the ghost mode in the spin connection, however the ghost appears in (classical) gravitational backreaction. The term linear in the field strength plays an important role in that construction; repeating their calculation  verbatim without the linear term results solely in trivial solutions $A=constant.$ However, working to higher orders in torsion or relaxing linearization assumption may illuminate interesting classical features of (\ref{Lagrangian}).

There is also a connection between the model considered in this work and the so-called $BF$ formulation of General Relativity \cite{Krasnov:2017epi,Krasnov:2021zen,Celada:2016jdt}. That story begins with an $SL(2,\mathbb{C})$ action written down by Pleb\'anski in \cite{Plebanski:1977zz}, which is a genuine gauge theory without \textit{a priori} knowledge of the spacetime metric. The field content includes  the connection's curvature $F$ (appearing at first order in the Lagrangian), an additional 2-form field $B$, and Lagrange multipliers. General Relativity is recovered by invoking appropriate reality conditions before identifying the spacetime metric as a nontrivial contraction of three copies of the $B$-field. Yang-Mills theory can be cast in a very similar manner in the $BF$ formalism; starting with $\mathcal{L}\sim B_A\wedge F^A+g^2B_A\wedge *B^A$ and integrating out $B_A$ leaves $\mathcal{L}\sim \frac{1}{g^2}\big( *F_A\wedge F^A\big).$ The gravity theory quadratic in the spin connection's field strength likely can be obtained by an analogous procedure. %It would be interesting to understand if the gravitational theory can be (uv?)extended by making use of an auxiliary field in analogy to the Yang-Mills case, resulting in a pure $F^2$ theory of gravity similar to what we have studied here.

We have additionally discussed various approaches towards taming the potential pathologies associated with a non-compact gauge theory. The essential conclusion from each approach - BRST, dynamical Killing form, or gauge fixing - is that the healthy states transform in the maximal compact subgroup $O(3)\subset SO(3,1)$. A natural setting for this to be realized without explicitly imposing a gauge fixing choice is a universe is filled with a ghost condensate \cite{Arkani-Hamed:2003pdi}.

At loop-level, the non-compactness of $SO(3,1)$ presents a subtle difficulty due to the fact that any unitary cut necessitates the spin-connection be well defined as an asymptotically free field, as we discuss in section \ref{sec:noncompactness}. The mass dimension of the current-current interaction is $d=6$ and since we considered 3+1 dimensional spacetime, the derived operator is irrelevant in agreement with the expectation that our result is only valid in the IR. This is also in analogy with the usual four-Fermi interaction in the electroweak theory.  

Throughout this paper, we did not specify the properties of the fermion field beyond assuming it is a Dirac spinor. We can therefore consider a few different scenarios that we will pursue in upcoming work. First, it is possible that the spin connection interacts universally with all fermions in the Standard Model, as one may expect from pure General Relativity. This would imply a new right-right interaction between standard model fermions that has not been observed.  It is interesting to entertain identifying such right handed term as a right handed Sterile Neutrino.

Alternatively, if the spin connection is blind to the Standard Model, the fermions considered here could be from a dark sector. The chiral properties of the dark fermions would naturally differ from the Standard Model in that case. Finally, we can consider the exciting possibility that the right-handed interaction corresponds to the dark sector, while the left-handed interaction is that of the Standard Model weak force. This would be an elegant realization of not only gravi-weak unification \cite{Alexander:2012ge}, but a `natural' dark matter candidate as well, with properties uniquely distinguishable from the Standard Model. There are various other approaches to relating the Standard Model to gravity, such as the non-commutative approaches of \cite{Connes:2006qv,Chamseddine:2006ep,Aydemir:2014ama,Aydemir:2015nfa,Aydemir:2016xtj,Aydemir:2018cbb}. The non-commutative geometrical approach has the virtue of preserving only universal terms in the induced action following the process of integrating out gravitational degrees of freedom \cite{Connes:2006qv,Chamseddine:2006ep}, where as our result leaves a non-universal operator. However, as is common in unified theories, the non-commutative approach necessitates enlarging the gauge group while we were able to focus solely on the Lorentz group. 

From the perspective or Effective Field theory, it is essential for the Yang-Mills theory discussed in this work to be directly connected to classical gravity. The simplest way to proceed would be to include the term linear in $F_{\mu\nu}^{ab}$, which corresponds to the Ricci scalar after invoking the metricity constraint. A possible route towards preserving the physics of the fermion-spin connection interaction is to decompose the connection into a background piece plus a (quantum) perturbation, $A=\bar{A}+A_Q.$ The classical part $\bar{A}$ would then be determined by the metricity constraint, while $A_Q$ would play a role in the interacting quantum theory. Inserting the decomposition into the theory $\mathcal{L}\sim F-\frac{1}{4}F^2$ results in terms that are beyond what we have analyzed in this work, with potential consequences for classical observables.

\iffalse 
There are numerous additional avenues to explore: potential phenomenology: cosmology? Dark matter? Dark energy ? External legs of spin connections (perhaps relevant for early universe physics, in analogy with quark-gluon plasma), further properties of composite states bound by spin connections (I think loop considerations are relevant here, so we would need to make sense of the asymptotic states in light of the unitarity cut).
\fi

%=================================================================================================================================
%=================================================================================================================================

\vspace{5mm}

%=================================================================================================================================
%=================================================================================================================================

\section*{Acknowledgements} 
The authors thank Adam Ball, Matthew Baumgart, Cliff Burgess, Humberto Gilmer, Mark Hertzberg, and Luke Lippstreu for helpful discussions while this work was in progress. We also thank John Donoghue, Joao Magueijo, and Tanmay Vachaspati for useful comments on an early draft. SA and TM are supported by the Simons Foundation, Award 896696.

\bibliographystyle{hunsrt}
\bibliography{refs2}

\iffalse

\newpage

\section*{On classical cosmological considerations}

Per the idea floated last Thursday, we can imagine that the spin connection perhaps has a nonzero (classical) expectation value in a cosmological background, inspired by \cite{Arkani-Hamed:2003pdi}. Immediate comments: assuming the components of the spin connection are constant, we appear to have exact solutions that are such that the scale factor is linear in time, $a(t)\sim t$, which is of potential interest in pre-inflationary physics. Said solution was obtained under some reasonable assumptions that I can spell out in person.

We may be able to think about this setup as being a theory such as 
$$\mathcal{L}\sim R(g)-\frac{1}{4}F_{\mu\nu}^{ab}F^{\mu\nu}_{ab},$$ 
and take 
$$A^{ab}_{\mu}=\bar{A}^{ab}_{\mu}+\delta A^{ab}_\mu,$$
where $\bar{A}$ is the classical matter source and $\delta A$ is a (quantum) fluctuation, whose properties are discussed in this note.
\fi

\end{document}